\documentclass[fdpMODIFIED,a4paper,fleqn%
]{w-artMODIFIED}
\usepackage{times,cite,w-thm}
\theoremstyle{plain}

\theoremstyle{definition}

\usepackage[]{graphicx}
\begin{document}
\DOIsuffix{theDOIsuffix}
\pagespan{1}{}



\title[AdS/CFT and Massive Gravity]{Multi-String Theories, Massive Gravity and the AdS/CFT\\ Correspondence}


\author[V. Niarchos]{Vasilis Niarchos\inst{1,}%
  \footnote{E-mail:~\textsf{niarchos@cpht.polytechnique.fr},
            }}
\address[\inst{1}]{Centre de Physique Th\'eorique, \'Ecole Polytechnique,
91128 Palaiseau, France\\
Unit\'e mixte de Recherche 7644, CNRS}
\begin{abstract}
  Infrared modifications of gravity have been proposed over the years as potentially
  useful phenomenological models. Massive (multi)-gravity is an interesting class of
  such theories. The true nature and the ultimate consistency of many of these models
  remains, however, unclear as one is frequently faced with important problems, $e.g.$ 
  classical instabilities and strong coupling problems which are hard to resolve without 
  a known UV completion of the theory. In this note, we review a recent attempt to study 
  some of these well known problems with the use of the AdS/CFT correspondence. 
  In this context, products of large-$N$ conformal field theories coupled by multi-trace
  interactions in diverse dimensions are used to define quantum multi-gravity 
  (multi-string/M theory) on a union of (asymptotically) AdS spaces. One-loop effects 
  generate a small $O(1/N)$ mass for the gravitons and provide non-trivial 
  examples of massive multi-graviton theories. 
  \end{abstract}
\maketitle                   






\section{Introduction}
\label{sec:intro}

One of the most vexing problems in theoretical physics is the cosmological
constant problem -- currently we observe that the universe accelerates at a
rate compatible with a vacuum energy density approximately equal to 
$(10^{-27}~M_P)^4$. This is a fundamental problem because it appears to 
challenge our ideas about the laws of nature simultaneously at long and short 
distance scales. 

A popular approach to this problem is based on the idea that gravity should be 
modified in the infrared (IR) in an attempt to effectively screen the gravitational 
interaction. Since the Einstein term in gravity is a two-derivative term, any local
IR modification of gravity proceeds via a potential for the graviton. A quadratic
mass term is the leading non-trivial part of the potential, but higher order terms 
can be present. Giving a mass to the graviton is a possibility that was explored
many years ago by Fierz and Pauli \cite{fp}. Later it was realized, following the 
important work of S.\ Deser and collaborators, that gravitational theories with a 
mass term are behaving quite unlike other theories and are typically accompanied 
by several problems, most notably ghost/tachyon instabilities and strong 
coupling issues.

At the same time, discussions of massive gravity in the context of the cosmological 
constant problem have been inconclusive. Using the cosmological equations for 
massive gravity \cite{gb} one finds that there is an effective cosmological acceleration 
at late times, which is equivalent to a constant {\it positive} vacuum energy 
$\Lambda^4\sim m_g^2 M_P^2$ \cite{review}. If the mass of the graviton is of order the 
inverse Hubble scale today, $m_g \sim H_0^{-1}$, then $\Lambda \sim 10^{-3} eV$ in 
agreement with todays' observations. This result is encouraging, but it turns out
that higher terms in the graviton potential give more and more dominant contributions
to the late time evolution of the universe \cite{review} indicating that we need a much 
better understanding of the theory. 

Our purpose in this short review of Ref.\ \cite{kirniar1} is to discuss the fundamental 
issues of massive graviton theories in a context where an alternative description of the 
theory provides us with a much needed additional control. Before going into the details 
of this context let us briefly summarize the main features of the fundamental issues that 
we want to understand better.

It has been shown in \cite{deser1,deser2,deser3,deser4}, and more recently in 
\cite{insta1,insta2,insta3}, that massive graviton theories are generically unstable. 
The unstable mode (usually called a Boulware-Deser mode) is a ghost ($i.e.$ a 
mode with a wrong sign kinetic term), and sometimes also a tachyon (a mode with 
a negative mass squared). The source of the instability lies in the non-linear structure 
of the kinetic terms. Although it is possible sometimes to fine-tune the instabilities away 
in a {\it fixed background}, small departures from the original background are enough
to reinstate them. Even in cases where one considers non-Lorentz invariant potentials
as in \cite{ghost}, and other cases reviewed in \cite{rreview}, the theory is stable only 
in a subclass of backgrounds. This behavior persists also in more exotic cases where 
the massive graviton is a resonance ($e.g.$ in brane-induced gravity \cite{greg}).

Another peculiar feature of massive gravity is the appearance of a strong coupling
scale at energies that are hierarchically smaller than the Planck scale. To be precise,
if the graviton mass $m_g \ll M_P$, then the theory becomes strongly coupled at
energies comparable or larger than $\Lambda_{\rm V}=(m_g^4M_P)^{\frac{1}{5}}$ 
\cite{vain}, a scale that is hierarchically smaller than the Planck scale. The interactions 
that become strongest are those of the scalar mode of the graviton. By judiciously 
modifying the graviton potential \cite{ags} the scale where gravity becomes strong can 
be improved to $\Lambda_{\rm AGS}=(m_g^2 M_P)^{\frac{1}{3}}$. 

Thinking of the massive graviton actions as low-energy effective actions, the above 
scales should be viewed as UV cutoffs in the standard sense of quantum field theory.
Beyond the energy of these scales one must know the UV completion of the theory in
order to proceed. For massive graviton theories this UV completion is essentially 
unknown and this poses a major hurdle in making further progress. This peculiar IR
behavior of massive gravity is reflecting an important feature: the failure of IR physics
to decouple in a simple way from UV physics. The theory has two characteristic scales,
$m_g$ and $M_P$, however, there is a dynamically generated intermediate scale where
the effective low-energy description breaks down. 

We should point out that neither $\Lambda_{\rm V}$ nor the improved $\Lambda_{\rm AGS}$ 
are cutoffs compatible with todays' experimental results on the gravitational force. Only a 
cutoff that would scale as $\Lambda_* \sim \sqrt{m_g M_P}$ would be marginally 
compatible with experiment, but a Lorentz-invariant massive graviton theory with 
such a cutoff is currently unknown (nevertheless some examples in AdS backgrounds 
were presented in \cite{kirniar1}).

Finally, it will be useful to introduce a related concept, that of multi-graviton theories.
To formulate a massive gravity theory, or to write down a general potential for the
graviton, a fiducial metric must be introduced. This metric is non-dynamical and is
typically taken to be flat (constant curvature metrics are also considered). A natural
generalization involves making the fiducial metric dynamical. In that case, we are
dealing with a bi-gravity theory\footnote{Bi-gravity was first introduced in 
\cite{salam1,salam2} in the context of the strong interactions. For a recent detailed 
classical study of such theories and their solutions see \cite{kd1,kd2,kd3}.} -- a 
theory with two propagating gravitons, two in general distinct Planck scales and 
in the decoupling limit two diffeomorphism invariances. By adding a potential that 
couples the two gravitons together it is possible to break one of the diffeomorphism 
invariances and give mass to a linear combination of the two gravitons. Bi-gravity 
theories naturally generalize to multi-gravity theories involving more than two metrics. 
We will soon encounter such theories in the context string theory.

\section{Conformal Field Theories and Multi-gravity in AdS spaces}
\label{sec:field}

In the presence of rather low UV cutoffs, it is sensible to seek out theories of
modified gravity which are UV complete. The AdS/CFT correspondence provides
a rather rare context where a theory of gravity (in asymptotically AdS spaces) has
an implicit UV completion. The completion is provided by a lower dimensional 
quantum field theory without gravity living on the boundary of the AdS space. 
Examples where the gravity on the AdS side is massive are an interesting arena 
where the problems of massive gravity outlined above can be discussed from a 
totally new and different perspective: the perspective of the non-gravitational field 
theory on the boundary.

A natural set of AdS/CFT examples with massive gravity can be obtained following
a rather simple idea. In standard AdS/CFT pairs the graviton in AdS maps to the 
stress-energy tensor of the conformal field theory (CFT). Accordingly, diffeomorphism 
invariance in AdS maps to energy-momentum conservation in the CFT. Deforming the 
CFT in a way that violates the energy-momentum conservation corresponds in AdS to a
gravitational Higgs effect where diffeomorphism invariance gets broken and the 
graviton acquires a mass.

There are several ways to violate the stress-energy conservation on the boundary.
One of them involves the introduction of $(d-1)$-dimensional defects \cite{kra}. 
Another involves product conformal field theories deformed by multi-trace deformations
\cite{kir,ack}. In what follows we will discuss massive (multi)-gravity in the second setup.

In order to be concrete, we will consider the following instructive prototype (several
generalizations can be found in \cite{kirniar1}). Our 
setup consists of two $d$-dimensional large-$N$ CFTs (CFT$_1$ and CFT$_2$)
which are coupled by a double-trace interaction $g{\cal O}_1 {\cal O}_2$, where 
${\cal O}_{1,2}$ are single-trace (scalar) operators in CFT$_{1,2}$ respectively. 
The two CFTs are generic -- we will only demand that they are defined on the same 
space and that each of them has a dual description in terms of a gravity (string/M) theory 
on a space of the form AdS$_{d+1}\times {\cal M}_i$, where ${\cal M}_i$ $(i=1,2)$ 
is some compact space. Coupling the CFTs by a multiple trace deformation is the 
generic way to make the CFTs communicate without spoiling their individual gauge 
symmetries. In order to preserve the standard large-$N$ counting the double-trace 
coupling constant $g$ must scale with $N$ as ${\cal O}(N^0)$.

Without the double-trace interaction the field theory on the boundary is the product
of two decoupled CFTs. This theory maps to a product of two string/M theories on
a product manifold of the form $\prod_i$ AdS$_{d+1}\times {\cal M}_i$ \cite{kir,ack}. 
In cases, where the (super)gravity approximation is valid this theory reduces to a 
bi-(super)gravity theory. This is a trivial bi-gravity theory in the sense that there are 
two gravitons that do not communicate with each other. 

The presence of the double-trace coupling $g$ on the boundary makes the dual 
bi-gravity theory a far more interesting system. The two CFTs are now coupled 
non-trivially. The total stress-energy tensor
\begin{equation}
\label{aaa}
T^{\rm tot}_{\mu\nu}=T_{\mu\nu}^{(1)}+T_{\mu\nu}^{(2)}-g\eta_{\mu\nu}
{\cal O}_1 {\cal O}_2
\end{equation}
remains conserved. An orthogonal combination of the stress-energy tensors,
however, is no longer conserved. Roughly speaking (for a more precise
expression see \cite{ack})
\begin{equation}
\label{aab}
\partial^\mu(T^{(1)}_{\mu\nu}-T^{(2)}_{\mu\nu})
\sim g \left[ (\partial_\nu {\cal O}_1){\cal O}_2-{\cal O}_1(\partial_\nu {\cal O}_2)\right]
~.
\end{equation}

In the bulk product of AdS spaces a corresponding breaking of a subgroup of the 
diffeomorphism symmetries occurs. Simultaneously, a linear combination of the 
gravitons becomes massive. Field theory predicts \cite{kir,ack} a mass determined 
by the scaling dimension of the non-conserved stress-energy tensor 
\begin{equation}
\label{aac}
m_g^2=d(\Delta_T-d) \sim \frac{g^2}{N^2}
~.
\end{equation}
In gravity this mass comes about in the following way.

The effect of multiple trace deformations in the AdS/CFT correspondence was analyzed
at tree-level ($i.e.$ to leading order in the $1/N$ expansion) some time ago in Refs. 
\cite{witten,berkooz,muck,minces,petkou}.
In situations where we can trust the supergravity approximation, the main effect of
field theory multi-trace deformations in the bulk is a change of the standard Dirichlet
or Neumann boundary conditions to mixed boundary conditions for the (scalar) fields  
$\phi_i$ that are dual to the operators ${\cal O}_i$ participating in the deformation.

\begin{figure}[t!]
\centering
\includegraphics[width=5cm]{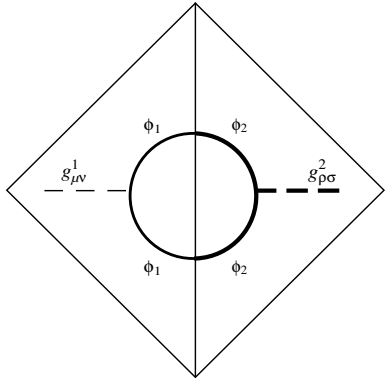}
\caption{A diagrammatic representation of the one-loop amplitude that gives a mass
to a linear combination of two gravitons, each living on a different AdS space. The mixed
boundary conditions for the dual scalars $\phi_1$, $\phi_2$ give a non-vanishing 1-2
graviton propagator.}
\label{loopgraviton}
\end{figure}

References \cite{kir,ack} (see \cite{porrati1,porrati2} for related earlier work) observed 
that the mixed boundary conditions facilitate a scalar loop correction to the graviton 
propagator in the bulk of the form depicted in Fig.\ \ref{loopgraviton}. This correction 
induces a non-zero mass term for a linear combination of the gravitons. The mass is 
of order $g/N$ as anticipated from the field theory on the boundary. A detailed 
computation of this mass can be found in \cite{ack} (a discussion of the graviton mass 
matrix in a more general context of a network of interacting CFTs can be found in 
appendix D of \cite{kirniar1}).

Further properties of this system can be read off the dual gauge theory. The double-trace
deformation can be relevant, marginal or irrelevant depending on the scaling dimensions
$\Delta_1, \Delta_2$ of the operators ${\cal O}_1, {\cal O}_2$. In general, the double-trace
deformation breaks the original conformal invariance and the couplings of the gauge 
theory run in the standard renormalization group (RG) sense. We would like to know the 
possible endpoints of these flows.

To uncover the full structure of the RG running in the space of multi-trace couplings
consider the general deformation
\begin{equation}
\label{aad}
\delta{\cal S}=\int d^d x\left[ N\sum_{i=1,2} g_i {\cal O}_i +
\sum_{i,j=1,2} g_{ij} {\cal O}_i {\cal O}_j+
\frac{1}{N} \sum_{i,j,k=1,2} g_{ijk} {\cal O}_i{\cal O}_j {\cal O}_k +
\cdots \right]
~,
\end{equation}
where the dots indicate higher multi-trace interactions. It will be convenient to denote
collectively the single-trace, double-trace, $etc$ couplings as
\begin{equation}
\label{aae}
g_{(1)}=\{ g_1,g_2 \}~, ~ ~ 
g_{(2)}=\{ g_{11},g_{12}, g_{22} \}
~, \cdots
\end{equation}
We will assume that the scaling dimensions $\Delta_1, \Delta_2 <\frac{d}{2}$ so that
the double-trace couplings are relevant. We would like to determine the RG flows and
the fixed points associated to the one-loop $\beta$-functions
\begin{equation}
\label{aaf}
\beta_i=\frac{\partial g_i}{\partial t}~, ~~
\beta_{ij}=\frac{\partial g_{ij}}{\partial t}~, ~~
\beta_{ijk}=\frac{\partial g_{ijk}}{\partial t}~, ~ ~\cdots
\end{equation}
where $t$ is the logarithm of the RG scale.

Performing an expansion in conformal perturbation theory up to second order in the
couplings one finds the following type of equations
\begin{subequations}
\begin{eqnarray}
\label{imrgaca}
\dot g_{(1)}&=&(d-\Delta_{(1)})g_{(1)}+C_{(1)(1)(1)} g_{(1)}^2
+C_{(1)(1)(2)} g_{(1)}g_{(2)}+
\\
&&+N^{-2}\left[
C_{(1)(1)(3)} g_{(1)}g_{(3)}
+ C_{(1)(1)(4)} g_{(1)}g_{(4)}
+C_{(1)(2)(2)} g_{(2)}^2
+C_{(1)(2)(3)} g_{(2)}g_{(3)}
\right]+\cdots
~,\nonumber
\end{eqnarray}
\vspace{-.5cm}
\begin{eqnarray}
\label{imrgacb}
\dot g_{(2)}&=&(d-\Delta_{(2)})g_{(2)}+
\\
&&+C_{(2)(1)(1)} g_{(1)}^2
+C_{(2)(1)(2)} g_{(1)}g_{(2)}
+C_{(2)(1)(3)} g_{(1)}g_{(3)}
+C_{(2)(2)(2)} g_{(2)}^2+
\nonumber\\
&&+N^{-2}\left[
C_{(2)(1)(4)} g_{(1)}g_{(4)}
+C_{(2)(2)(3)} g_{(2)}g_{(3)}
+C_{(2)(2)(4)} g_{(2)}g_{(4)}
+C_{(2)(3)(3)} g_{(3)}^2
\right] +\cdots
~, \nonumber
\end{eqnarray}
\vspace{-.5cm}
\begin{eqnarray}
\label{imrgacc}
\dot g_{(3)}&=&(d-\Delta_{(3)})g_{(3)}
+C_{(3)(1)(1)}g_{(1)}^2
+C_{(3)(1)(2)} g_{(1)}g_{(2)}+
\\
&&+C_{(3)(1)(3)} g_{(1)}g_{(3)}
+C_{(3)(1)(4)} g_{(1)}g_{(4)}
+C_{(3)(2)(2)} g_{(2)}^2
+C_{(3)(2)(3)} g_{(2)}g_{(3)}+
\nonumber\\
&&+N^{-2}\left[
C_{(3)(2)(4)} g_{(2)}g_{(4)}
+C_{(3)(3)(3)} g_{(3)}^2
+C_{(3)(3)(4)} g_{(3)}g_{(4)}
\right]+\cdots
~, ~ \cdots \nonumber
\end{eqnarray}
\end{subequations}
We included terms up to four-trace couplings in the above expressions
hoping that the general structure is evident. $C_{(m)(n)(p)}$ is shorthand 
for a set of 3-point function coefficients normalized in such a way that their
leading contribution in the $1/N$ expansion is of order $N^0$. More details 
can be found in \cite{kirniar1}. In what follows we would like to bring forward 
some of the key properties  implied by these equations.

First, to leading order in $1/N$ we can consistently set all the single-trace
couplings $g_{(1)}$ to zero at all scales. In search of fixed points, we observe 
that $\beta_{(2)}$ depend only $g_{(2)}$ and $\beta_{(i)}=0$, $i\geq 3$, determine
straightforwardly, in our order of approximation, the fixed point values of $g_{(i)}$,
$i\geq 3$, once the fixed point values of the double-trace couplings $g_{(2)}$ is 
computed. Hence, it is enough to solve the double-trace fixed point equations.
With appropriate normalization these equations, which are exact in the large-$N$
limit, read
\begin{subequations}
\begin{equation}
\label{bdrgca}
(d-2\Delta_1)g_{11}-8g^2_{11}-8g^2_{12}=0
~, 
\hspace{.4cm}
(d-2\Delta_2)g_{22}-8g^2_{22}-8g^2_{12}=0
~,
\end{equation}
\begin{equation}
\label{bdrgcc}
(d-\Delta_1-\Delta_2)g_{12}-8g_{12}(g_{11}+g_{22})=0
~.
\end{equation}
\end{subequations}
This is a simple set of quadratic equations with the following solutions. 

Non-interacting fixed points ($i.e.$ points with $g_{12}=0$) exist when
\begin{equation}
\label{rgac}
g_{11}(8g_{11}-(d-2\Delta_1))=0~, ~ ~ g_{22}(8g_{22}-(d-2\Delta_2))=0
~.
\end{equation}
Assuming $\Delta_1,\Delta_2\neq d/2$ these are points where the CFTs
1 and 2 are either undeformed ($g_{ii}=0$) or individually deformed to a
new fixed point with $g_{ii}=\frac{1}{8}(d-2\Delta_i)$. These theories  are 
well known \cite{witten}.

\begin{figure}[t!]
\centering
\includegraphics[width=5cm]{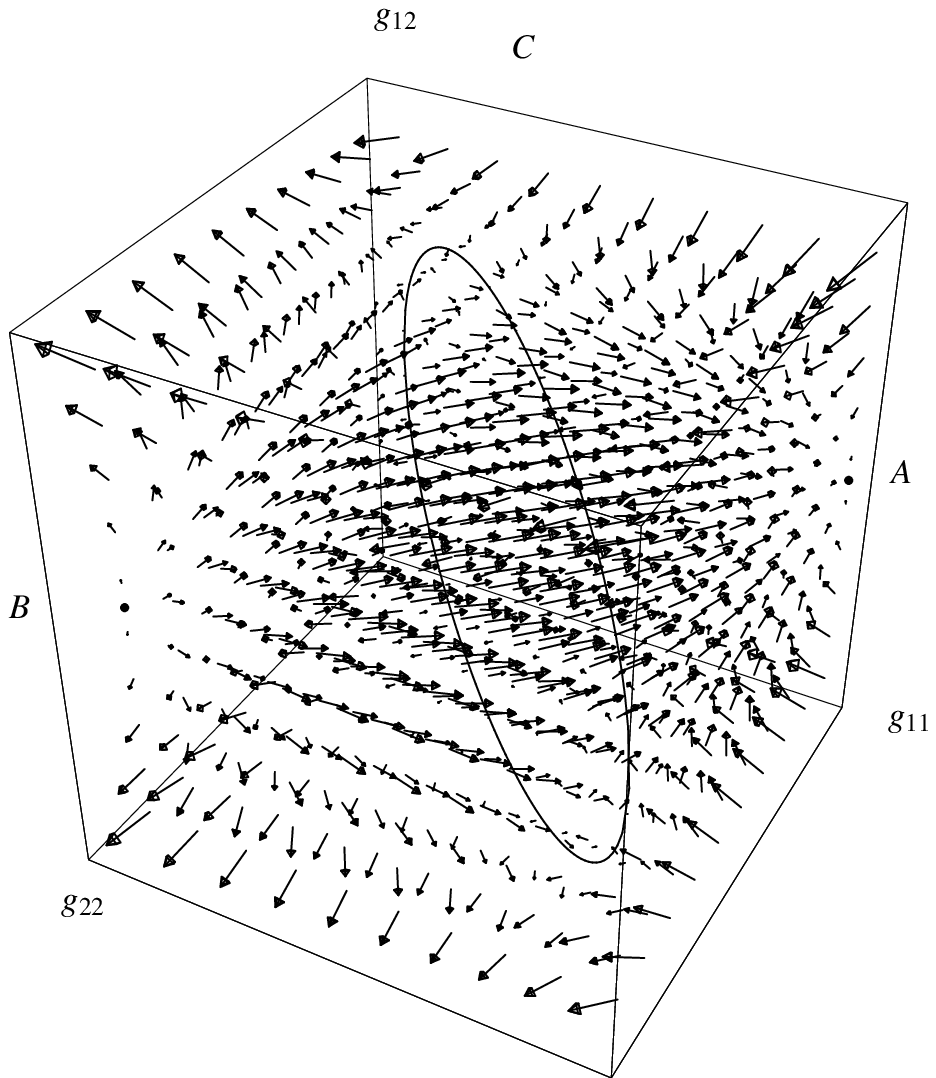}
\hspace{1cm}
\includegraphics[width=5cm]{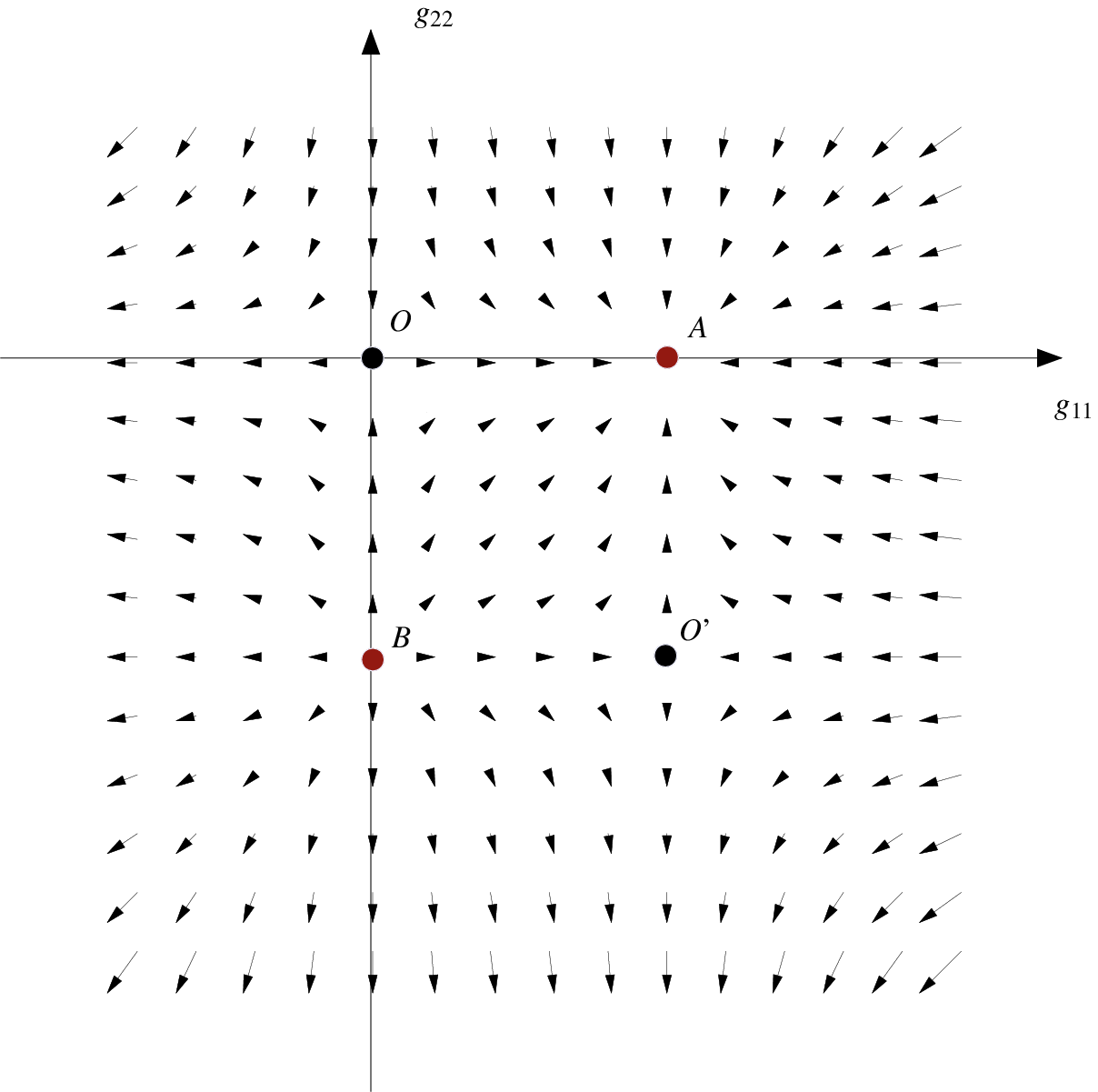}
\caption{The figure on the left is a depiction of the RG flows and fixed points
of the double-trace equations $\dot g_{(2)}=\beta_{(2)}$. The $\beta$-functions
vanish on the circle $C$ (at the center of the plot) and the special points $A$ and $B$
at the corners. The figure on the right is a 2D slice of the RG flow vector field along
the $g_{12}=0$ plane. The 3D plot on the left arises from the 2D plot on the right
by rotating the 2D plot in three dimensions around the $AB$ axis. $O$, $O'$ are the
points where the $C$ circle crosses the $g_{12}=0$ plane.}
\label{gfgraph}
\end{figure}

Interacting fixed points ($i.e.$ fixed points with $g_{12}\neq 0$) exist only when
$\Delta_1=\Delta_2$ or $\Delta_1+\Delta_2=d$ with $\Delta_1,\Delta_2\neq \frac{d}{2}$
in both cases. These points lie on a circle in the $(g_{11},g_{22},g_{12})$ space.
In a parametrization of the double-trace coupling space where $\Delta_1+\Delta_2=d$ 
at $g_{11}=g_{22}=g_{12}=0$ this circle -- denoted as circle $C$ in the left figure in Fig.\ 
\ref{gfgraph} -- is described by the equations
\begin{equation}
\label{rgad}
g_{11}=-g_{22}\equiv f~, ~ ~ g_{12}\equiv g~ ~ {\rm and} ~ ~
4f^2-af+4g^2=0~, ~ ~ 2a\equiv d-2\Delta_1=2\Delta_2-d
~.
\end{equation}
As is apparent from the right plot in Fig.\ \ref{gfgraph}, these points are 
repellors of the RG flow. When we perturb away from them, there are always
directions where the RG flow is repelling. There are two distinct
possibilities for the late RG time behavior of the system. Either it is driven 
towards the most attractive fixed point $A$, where $g_{12}=0$ and the interaction
between the CFTs is lost, or it is driven towards stronger and stronger coupling
where the $1/N$ expansion eventually breaks down.  

Incorporating the $1/N$ corrections does not modify this picture dramatically.
In perturbation theory, these corrections will slightly shift the fixed point values 
determined at leading order. An important new element should be noted however.
Already at the next-to-leading order there is a running of the single-trace couplings 
which is sourced by the evolution of the double-trace couplings.

\section{Lessons for Massive Gravity}

We can now ask what are the consequences of these properties in gauge theory 
for the dual massive multi-gravity theory on the AdS side of the correspondence. 
More importantly, what lessons can we extract for modified theories of gravity from 
this context?

In the setup we have just described, the dual gravitational theory is at tree-level
($i.e.$ to leading order in the $1/N$ expansion) a string/M-theory on a union of
AdS spaces. In the supergravity regime (whenever this is appropriate), the 
low-energy dynamics is described by a bi-gravity theory with mixed boundary 
conditions for a set of scalar fields. At this order, there are two {\it massless} 
non-interacting gravitons. Boundary RG flows do not manifest in the bulk as 
radial running of the background solution. This is consistent with the fact that 
in gauge theory there is no running of single-trace couplings at leading order 
in $1/N$.

The plot thickens beyond tree-level. First, one can argue on general grounds 
\cite{nonlocal} that the multi-trace deformed gauge theory defines a new type of
string theory, called non-local string theory (NLST), with non-standard features. 
This theory is non-local on the worldsheet -- one has to extend the 
perturbative expansion to include contact interactions between disconnected 
worldsheets. A clear manifestation of the rules of this new perturbative expansion 
has been given for 2D string theories in \cite{kirniar2}. NLSTs are also non-local in 
spacetime. Typically, the non-locality scale is set by the AdS scale \cite{Aharony:2005sh}.

Whatever the complicated structure of our multi-NLSTs is, we have seen that one-loop
effects generate a non-trivial graviton potential. One can imagine, in situations where
supergravity can be trusted, that the low-energy dynamics is described by a set of 
low-energy effective actions in the general spirit of bi-gravity theories of Damour
and Kogan \cite{kd1} (extra matter besides the gravitons has to be included in these 
actions). The tree-level AdS$\times$AdS background is not a solution of the equations 
of motion of this action. In agreement with the structure of RG flows in gauge theory,
the one-loop effects backreact and deform the tree-level background.

Gauge theory makes certain predictions for the structure of this new bi-gravity action.
First, the equations of motion of this action must have an interacting AdS$\times$AdS
product solution where some of the scalar fields have a non-trivial profile
and gravity is massive. These solutions map to the interacting products of CFTs
lying on the fixed point circle $C$ in Fig.\ \ref{gfgraph}. Furthermore, there should
be domain wall solutions interpolating between a `UV' AdS$\times$AdS geometry
with a massive graviton and an `IR' AdS$\times$AdS geometry with two massless
non-interacting gravitons. These solutions would correspond to the gauge theory 
RG flows from the fixed circle $C$ to the attractive point $A$. One could think of them 
as a dynamical way of switching off the graviton mass. In AdS space there is no vDVZ 
discontinuity \cite{vvd,porrativvd} which could prevent the existence of such solutions.

It is interesting to ask if we can use the gauge theory to learn anything about the typical 
problems that plague the massive graviton theories -- instabilities and strong coupling
issues. A naive calculation of strong coupling scales in massive gravity around AdS
(repeating the arguments of \cite{ags}) shows \cite{kirniar1} that the graviton interactions 
become strong at a scale set by the AdS scale. This scale appears, however, to be a fake. 
There are several precedents of such fake scales -- Kaluza-Klein gravity and string theory 
are such examples. In these cases, it is important to consider all the degrees of freedom
in the system and the calculation that leads to the above strong coupling scales around
the AdS scale does not take into account the interactions of the graviton with the other 
matter fields in the system.

There are several reasons why one should expect the absence of a strong coupling
problem in our massive graviton theories. At the AdS scale, which is the dangerous
scale for our naive estimates, new degrees of freedom appear as bound states of 
the dual scalars (these are crucial for the Higgs phenomenon in gravity). The AdS
scale is also a scale where our multi-string theories exhibit non-locality. At the same
time, on the dual gauge theory side there is no indication for an intermediate strong 
coupling scale.

The usual ghost/tachyon instabilities also appear to be absent in our examples. On the 
gauge theory side there is no sign of an instability which one would interpret as a 
Boulware-Deser mode. This would fit nicely with the absence of intermediate strong 
coupling scales, because the two are, in general, related \cite{insta1}.

What we learned in this short review can be summarized in a few lines as follows.
There are UV complete multi-graviton theories with light gravitons -- masses of the 
order ${\cal O}(1/N)$ -- which have a dual description in terms of a deformed 
product of large-$N$ CFTs. In these theories
\begin{itemize}
\item When the diffeomorphism invariance breaking is small, there are no strong
coupling problems and no Boulware-Deser instabilities.
\item Stable massive graviton backgrounds are not generic, they require fine-tuning.
\item Without fine-tuning the spacetime background backreacts either to `destroy' the
theory or to erase the effects of the graviton mass and trivialize the bi-gravity interactions.
The `destruction' of the theory occurs with a breakdown of the large-$N$ expansion or 
in 2D examples \cite{kirniar2} with a breakdown of the double-scaling limit that defines it. 
\end{itemize}

\begin{acknowledgement}
  I would like to thank Elias Kiritsis for an enjoyable collaboration on 
  this subject. I am also grateful to the organizers of the 2008 RTN meeting on
  ``Constituents, Fundamental Forces and Symmetries of the Universe'' in 
  Varna, Bulgaria, where I was given the opportunity to present this work. This 
  work has been supported by the European Union through an Individual Marie 
  Curie Intra-European Fellowship. 
  Additional support was provided by the ANR grant, ANR-05-BLAN-0079-02, 
  the RTN contracts MRTN-CT-2004-005104 and MRTN-CT-2004-503369, 
  and the CNRS PICS {\#} 3059, 3747 and 4172. 
\end{acknowledgement}

%
%

\end{document}